\documentclass[%
English,
preprint,
prl,
superscriptaddress,
 amsmath,amssymb,
 aps,
]{revtex4-2}
\usepackage{amsmath}
\usepackage{comment}
\usepackage{array}
\usepackage[table]{xcolor}
\usepackage[load-configurations=abbreviations]{siunitx}
\usepackage{appendix}
\usepackage{float}
\usepackage{graphicx}
\usepackage{dcolumn}
\usepackage{bm}
\usepackage{xcolor}

\begin{document}

\title{Air Drag Controls the Finite-Time Singularity of Euler’s Disk}

\author{Benjamin Thorne}
\email{bthorne@g.harvard.edu}
\author{Ahmad Zareei}%
\author{Kausalya Mahadevan}
\affiliation{School of Engineering and Applied Sciences, Harvard University, Cambridge, Massachusetts, USA}
\author{Shmuel M. Rubinstein}
\affiliation{School of Engineering and Applied Sciences, Harvard University, Cambridge, Massachusetts, USA}
\affiliation{The Racah Institute of Physics, The Hebrew University of Jerusalem, Jerusalem, Israel}
\author{Ariel Amir}
\affiliation{Department of Physics of Complex Systems, Weizmann Institute of Science, Rehovot, Israel}

\date{\today}

\begin{abstract}
The motion of a disk spinning to rest after being tipped on its side is a classic example of a finite-time singularity, yet the dominant dissipation mechanism governing this process remains debated. Using stereoscopic high-speed imaging, we study the dynamics of disks with varying mass and radius on different surfaces. We show that the late-time motion near the singularity is governed by viscous air-drag arising from shear in the boundary layer beneath the disk, as evidenced by the mass dependence of the dynamics, measurements in a partial vacuum, and a geometric control using a steel ring. At earlier times, dissipation is dominated by rolling friction, which on glass exhibits an unexpected sublinear scaling with disk mass, suggesting an adhesion-based rolling resistance. These results clarify the dissipation mechanisms underlying the singularity of Euler's disk and have broader implications for rolling-contact systems operating under low loads on smooth surfaces.
\end{abstract}

\maketitle

When a coin is spun on a table, it undergoes a short-lived but distinctive motion. It first rolls on its rim about its center of mass, the frequency rising rapidly as energy is lost, until finally it clatters to an abrupt halt. A popular physics toy known as \textit{Euler's Disk} (Toysmith, USA), is an optimized version of this system that consists of a heavy steel disk and a concave mirrored base. Its large mass and low-friction design allow it to spin for up to \SI{150}{\second}, and reach frequencies of over \SI{100}{\hertz}. Despite the simplicity of this system and its widespread popularity, the physical origin of its dramatic acceleration and sudden halt remains controversial.

In the frictionless limit, a disk rolling about its center of mass has long been understood. Conservation of angular momentum yields a relation between the precession frequency of the point of contact $\Omega$ and the tilt angle $\theta$:
\begin{equation}
    \Omega^2 \sin\theta = \frac{4g}{R},
    \label{adiabatic}
\end{equation}
where $R$ is the disk radius and $g$ is the gravitational acceleration \cite{olsson_coin_1972,petrie_does_2002}. The corresponding energy balance yields the dissipation rate 
\begin{equation}
    \dot{E} = \frac{3}{2} m g R \dot{\theta},
    \label{edot}
\end{equation}
where $m$ is the mass of the disk. Here, $1/3$ of the energy is kinetic, consistent with the virial theorem. The energy dissipation rate was first considered by Moffatt \cite{moffatt_eulers_2000} who demonstrated that if the dissipation rate $\Phi$, increases as $\theta\to0$, then the system approaches a so-called \textit{finite-time singularity}; this accounts for the apparent divergence of $\Omega$, even as the disk continues to lose energy. Moffatt went on to suggest that the energy dissipation is due to viscous air drag in the thin gap beneath the disk, $\Phi_{gap}=\pi\eta gR^2/\theta^2$, where $\eta$ is the air viscosity. Substitution into Eq. \ref{edot} and integrating yields a power-law behavior
\begin{equation}
    \Omega^{-2}\propto\theta = A(t_f-t)^n,
    \label{theta}
\end{equation}
where $A=(2\pi\eta R/m)^{1/3}$, $n=1/3$ and $t_f-t$ denotes the approximate time remaining before the disk stops.

Subsequent experiments confirmed the singular behavior and the general form of Eq. \ref{theta} but challenged the air-drag mechanism, finding the dynamics to be surface sensitive and the exponent larger than anticipated. Reported exponents include  $n=2/3$ \cite{easwar_speeding_2002,ma_rolling_2014}, $n=1/2$ \cite{mcdonald_rolling_2000,caps_rolling_2004}, or a crossover from $n=2/3$ at early times to $n=1/2$ at late times \cite{leine_experimental_2009}. The exponent $n=2/3$ is typically attributed to rolling friction with $\Phi \propto \Omega$ \cite{le_saux_dynamics_2005}, whereas $n=1/2$ implies $\Phi \propto \Omega^2$, consistent with viscous rolling friction \cite{leine_experimental_2009} or turbulent air drag \cite{bildsten_viscous_2002}. A further refinement of Moffatt’s theory accounting for viscous boundary-layer thickness yields $n=4/9$ \cite{bildsten_viscous_2002}, which is experimentally difficult to distinguish from $n=1/2$ \cite{darmendrail_euler_2021,collins_eulers_2022}. Hence, distinct dissipation mechanisms can produce nearly indistinguishable exponents, leaving the dominant mechanism unresolved. While $n$ only encodes the dependence of the energy dissipation on $\Omega$, the time-invariant pre-factor $A$ in Eq. \ref{theta} depends on physical parameters such as $m$ and $R$ that can be systematically varied. To our knowledge, no study has leveraged this dependence to identify the dominant dissipation mechanism.

Here we determine the dominant energy dissipation mechanism by systematically varying the disk mass and radius across different surfaces and using stereoscopic high-speed imaging to measure $\theta(t)$ and $\Omega(t)$.
We show that the late-time dynamics are consistent with dissipation due to air-drag in a viscous boundary layer beneath the disk, while the earlier stages of motion are governed by rolling friction. We verify the role of air drag through vacuum measurements, and a geometric control using a steel ring. In addition, we find that the rolling-friction-dominated dynamics at early times exhibit an unexpected sensitivity to disk mass, with heavier disks spinning significantly longer than lighter ones.

 \begin{figure}[!h]
\includegraphics[width=0.49\textwidth]{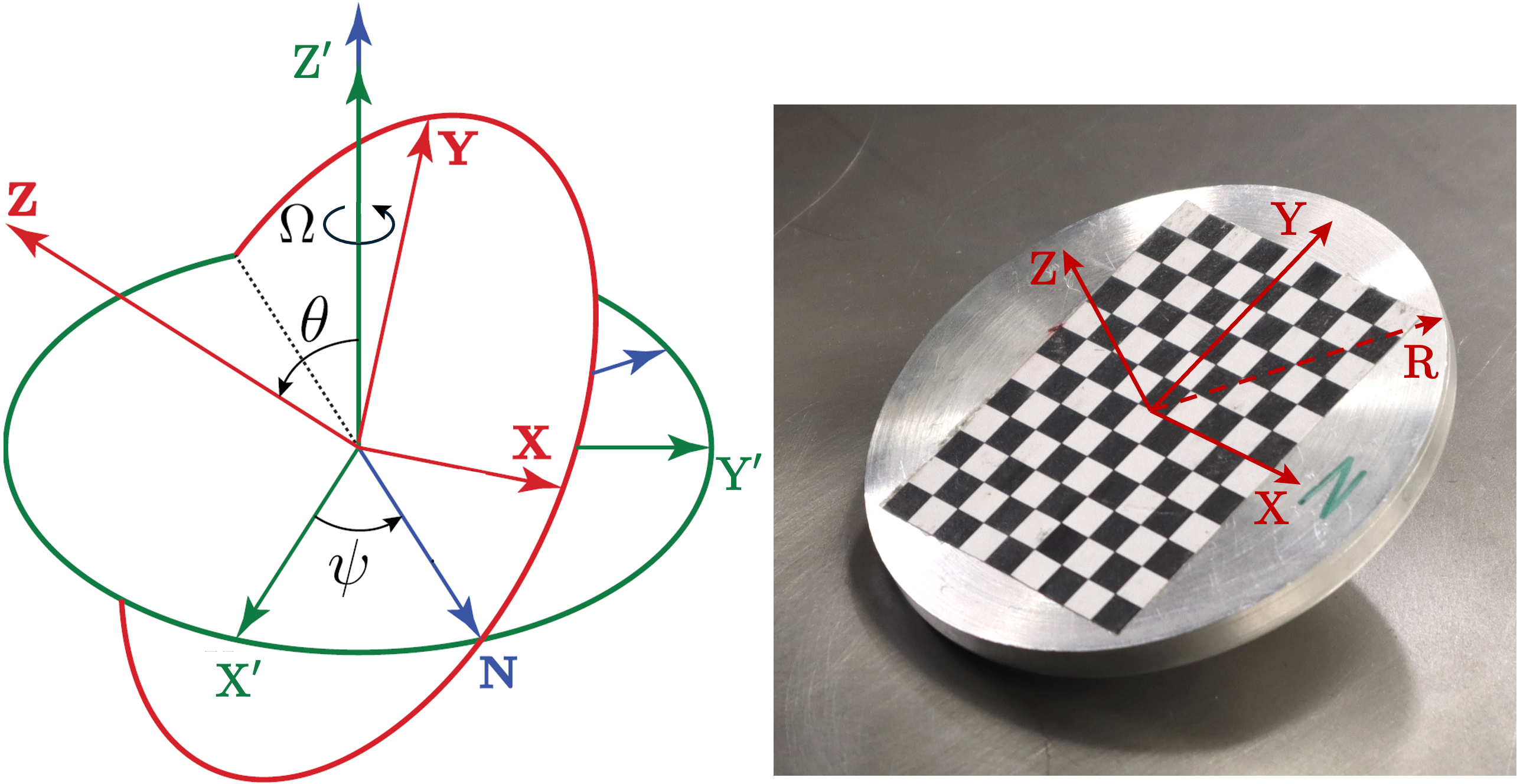}
\caption{\label{fig:sketch} \textbf{Left:} sketch of the Euler's disk system. The precession frequency $\Omega=\dot{\psi}$ corresponds to the frequency of the contact point. \textbf{Right:} $84$~g aluminum disk on steel surface. }
\end{figure}

We conduct our experiments by machining steel and aluminum disks with a range of thicknesses and diameters, with some disks featuring a rounded 1.6-mm-radius edge fillet while others are left sharp-edged; all disk parameters are summarized in Table S1. Test surfaces consist of flat 60-cm $\times$ 60-cm sheets of glass (3.2-mm thickness), aluminum (2.3-mm thickness), and steel (3.4-mm thickness), mounted on butyl rubber to reduce vibrations; all surfaces were used as received with a factory finish.

A $9 \times 12$ checkerboard pattern with 4.7~mm squares is printed on paper and attached to the top surface of each disk using double-sided tape; see Fig. \ref{fig:sketch}. A pair of synchronized high-speed cameras (Phantom, USA) record the final 10 s of motion at 1000 frames per second from different viewing angles. The three-dimensional positions of the corners of each checkerboard square are reconstructed using MATLAB's stereo camera calibration toolkit \cite{zhang_flexible_2000,ma_rolling_2014}. The points are then fit to a plane, from which the tilt angle $\theta$ and precession angle $\psi$ are extracted relative to the disk position at rest; the frequency $\Omega(t)$ is then obtained by differentiating $\psi(t)$. A sketch of the system, as well as a photograph of one of the disks are shown in Fig. \ref{fig:sketch}. Previous studies have shown that the late-time dynamics of Euler's disk are repeatable despite variability in initial conditions \cite{easwar_speeding_2002,leine_experimental_2009,caps_rolling_2004}; therefore all disks are spun by hand. 

We define $t_f$ as the extrapolated collapse time of the asymptotic power-law in Eq. \ref{theta}. Operationally, $t_f$ is chosen such that the final 0.1 s of the trajectory is maximally linear in log–log space (minimizing residuals). This definition typically places $t_f$ approximately 3 ms before the last tracked motion of the disk (Fig. S1 \cite{supplemental}). We interpret the short interval of motion after $t_f$ as a loss of contact between the disk and the surface, consistent with previous reports \cite{kessler_ringing_2002,borisov_loss_2015,collins_eulers_2022}.

\begin{figure}[!htb]
\includegraphics[width=.455\textwidth]{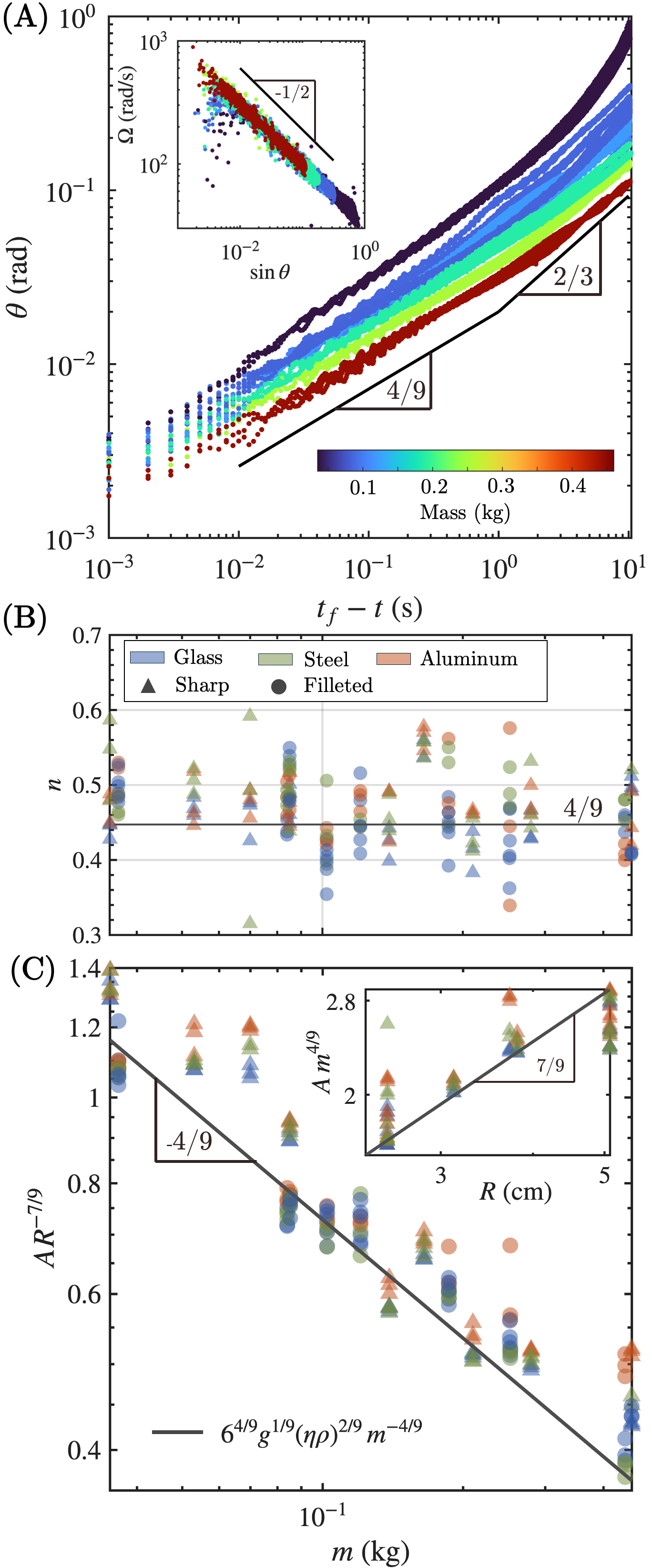}
\caption{\label{fig:glasstheta} \textbf{A)} $\theta$ vs. $t_f-t$ for filleted disks of varying mass and constant diameter on glass. Three repetitions per disk. Inset: Corresponding $\Omega$ vs. $\sin\theta$ data with best fit $\Omega\propto(\sin\theta)^{-.49}$. \textbf{B)} The exponent $n$ from Eq. \ref{theta} fitted to the final 0.1 s of motion for disks of varying mass for both sharp-edged and filleted disks on different surfaces. \textbf{C)} Prefactor $A$ from fits to Eq. \ref{theta} scaled by $R^{-7/9}$ for the last 0.1s of motion. Inset: $A$ vs. $R$ for sharp disks with varying radius; scaled by $m^{4/9}$.}
\end{figure}

Over the final 10 s of motion, two trends emerge in $\theta(t_f-t)$. First, the power-law exponent $n$ decreases as the disk approaches rest, suggesting a change in the dominant dissipation mechanism. Second, the data exhibit a pronounced dependence on disk mass, with heavier disks reaching smaller tilt angles than lighter ones at a given value of $t_f - t$, as shown in Fig.~\ref{fig:glasstheta}A. Throughout the motion, the disks remain in good rolling contact and satisfy Eq.~\ref{adiabatic}, as shown by the collapse of $\Omega(\sin\theta)$ for all masses in Fig. \ref{fig:glasstheta}A (inset). This also confirms that the tilt-angle and frequency exponents are related by $n_\Omega = -n_\theta/2$. Additional data for steel and aluminum surfaces, as well as sharp-edged disks are shown in Figs. S2-S3 \cite{supplemental}.

Typical rolling-friction models cannot account for the mass dependence observed in Fig. \ref{fig:glasstheta}A. This follows from the assumption that dissipation is proportional to normal force, $\Phi_{\mathrm{roll}}=\mu mgR\Omega$, where $\mu$ is a friction coefficient \cite{easwar_speeding_2002,le_saux_dynamics_2005,leine_experimental_2009,ma_rolling_2014}. Because the total mechanical energy is also proportional to mass, substitution into Eq. \ref{edot} predicts mass independent dynamics. Air-drag, by contrast, naturally introduces a mass dependence: the drag depends primarily on disk radius, while the total mechanical energy increases linearly with disk mass. Heavier disks are therefore expected to spin faster than lighter ones, analogous to the higher terminal speeds of denser objects falling through a fluid.

Motivated by this mass dependence, we consider viscous air drag as a candidate mechanism governing the late-time dynamics. Bildsten \cite{bildsten_viscous_2002} modeled the dissipation arising from shear in the viscous boundary layer beneath the disk as $\Phi_{\mathrm{BL}} = 4 g^{5/4} R^{11/4} \sqrt{\eta \rho}\, \theta^{-5/4},$ where $\rho$ is the air density. Substituting this into Eq. \ref{edot} and integrating yields
\begin{equation}
    \theta(t) =
    \frac{6^{4/9} g^{1/9} (\eta \rho)^{2/9} R^{7/9}}{m^{4/9}}
    (t_f - t)^{4/9}
    \label{bildsten}
\end{equation}
which accounts for the observed mass dependence, and predicts $n=4/9$.

Fitting the final $0.1$~s of data in Figs. \ref{fig:glasstheta}A and S2 to Eq. \ref{theta} yields an average decay exponent $n = 0.46 \pm 0.04$ across all disks, consistent with the boundary-layer prediction of Eq. \ref{bildsten}. While a weak trend toward smaller values of $n$ is observed for increasing disk mass, this variation is modest and may reflect residual contributions from dissipation at earlier times. A plot of the fitted exponents vs. $m$ is shown in Fig. \ref{fig:glasstheta}B. Fixing the exponent to $n=4/9$ and fitting $\theta(t)=A(t_f-t)^{4/9}$, the free parameter $A$ closely follows the prediction of Eq.~\ref{bildsten} without the use of adjustable parameters; see Fig.~\ref{fig:glasstheta}C. Because Eq.~\ref{bildsten} is derived from a scaling argument, the prefactor is not expected to be exact; confirming it precisely would require a full numerical solution of the coupled fluid–structure problem. Nevertheless, for sharp-edged disks of varying radius, we likewise observe consistency with the predicted scaling $A \propto R^{7/9}$; see Fig.~\ref{fig:glasstheta}C (inset). Moreover, the collapse of this data for steel, glass, and aluminum demonstrates surface insensitivity consistent with air-drag dissipation. By contrast, the early-time dynamics are strongly surface sensitive (Figs. S2 and S5A). 

Previous experimental studies have examined the role of air drag in Euler's disk by analyzing the motion of disks in a partial vacuum and concluded that its effect is negligible \cite{engh_numismatic_2000,easwar_speeding_2002}. These studies primarily focused on changes in the overall stopping time and implicitly assumed that air drag would dominate the dissipation throughout the motion. If, however, viscous air drag becomes significant only during the final stages of motion, its influence on the total stopping time may be too small to detect, particularly for disks with large $\mu$ \cite{engh_numismatic_2000}. Moreover, in Moffatt's air-drag model the dissipation rate depends only on the air viscosity and is therefore largely insensitive to pressure; consequently, no substantial difference would be expected under partial vacuum. The viscous boundary-layer model in Eq. \ref{bildsten} predicts that air drag should, conversely, exhibit a weak but measurable dependence on the air density as $\rho^{2/9}$.

\begin{figure}[!htb]
\includegraphics[width=0.48\textwidth]{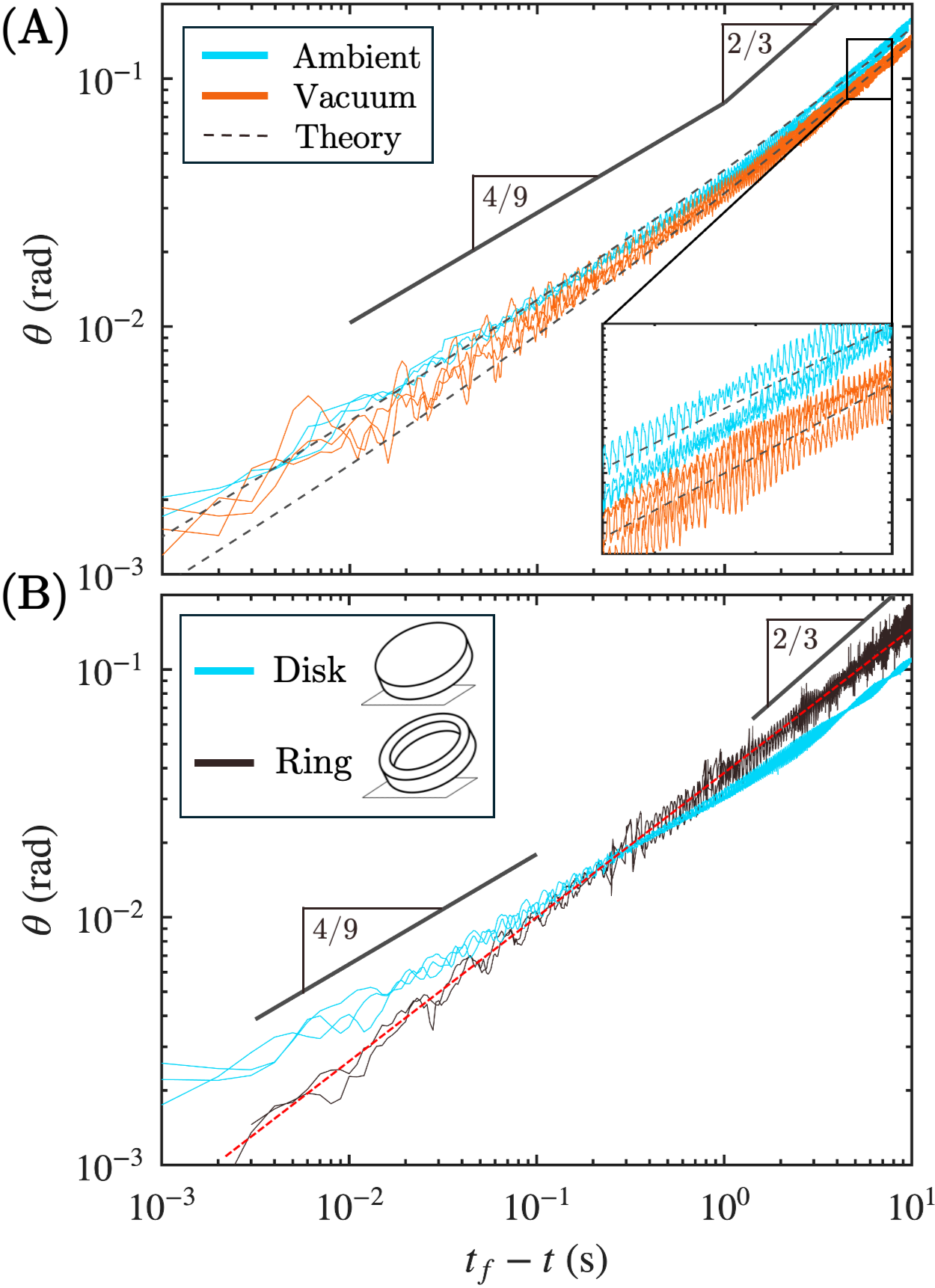}
\caption{\label{fig:vac} \textbf{A)} $\theta$ vs. $t_f-t$ for a 445 g steel disk on glass in ambient conditions and in a partial vacuum of 0.1 atm. Dashed lines are solutions to Eq. \ref{dragfric} with fitted $\mu=10^{-4}$,  as well as $\rho_{amb.}=1.18$ kg/m\textsuperscript{3} and $\rho_{vac}=0.118$ kg/m\textsuperscript{3}. Inset: Enlarged section. \textbf{B)} $\theta$ vs. $t_f-t$ on a glass plate for a 445 g steel disk and a 740 g steel annulus. A power-law fit is shown in red as a visual aid. Frequency ($\Omega$) data are shown in Fig. S4.}
\end{figure}

To test this prediction, we examine the late-time dynamics of disks spun on glass in a partial vacuum at 0.1~atm. Disks are spun by hand at atmospheric pressure, after which the chamber is sealed and evacuated over a period of roughly 30~s. The disk motion is recorded during the final 10~s using stereoscopic cameras positioned outside the chamber. Since the total spinning time is typically on the order of 120~s, the final portion of the motion occurs at effectively constant pressure. The control experiment at ambient pressure is performed in the same chamber with no applied vacuum. Under partial vacuum, we observe a reduction in $\theta(t_f - t)$, indicating reduced dissipation, as shown in Fig.~\ref{fig:vac}A. The magnitude of this reduction is consistent with Eq.~\ref{bildsten} when rolling friction is considered alongside viscous air drag.

The early-time dynamics are dominated by a surface-sensitive behavior with an effective exponent $n \approx 2/3$ which we attribute to rolling friction, as shown in Figs.~\ref{fig:glasstheta}A and S2 \cite{supplemental}. To model both rolling-friction and air-drag dissipation, we set $-\dot{E}=\Phi_{\mathrm{BL}}+\Phi_{\mathrm{roll}}$ yielding
\begin{equation}
    \frac{3}{2}mgR\dot{\theta}
    =
    -4g^{5/4}R^{11/4}\frac{\sqrt{\eta\rho}}{\theta^{5/4}}
    - \mu mg R \cos\theta \, \Omega.
    \label{dragfric}
\end{equation}
Numerical integration of Eq. \ref{dragfric} yields $\theta(t)$, which is used to generate the theoretical predictions shown in Fig.~\ref{fig:vac}A.

The solution exhibits a crossover from $n = 2/3$ at early times, when rolling friction dominates, to $n = 4/9$ at late times, when viscous air drag becomes significant. Because the air-drag term depends explicitly on the air density $\rho$, the solution becomes increasingly pressure sensitive at late-times. An estimate for the crossover angle can be obtained by equating the two dissipation rates, yielding
$\theta_{\mathrm{c}}=\left(\frac{2 R^{9/4} \sqrt{\eta \rho}}{\mu mg^{1/4}}\right)^{4/3}$. Fitting the data in Fig.~\ref{fig:vac}A at ambient pressure and using $\mu$ as a free parameter, yields $\theta_{\mathrm{c}} \approx 3 \times 10^{-2}$~rad, in reasonable agreement with the experimental data \footnote{This crossover is very sensitive to $\mu$ so slightly different data yield different results. For example, the 450~g disk in Fig. \ref{fig:glasstheta}A gives $\theta_c=5 \times 10^{-2}$~rad.}.

At sufficiently low pressures, an additional crossover is predicted due to the growth of the viscous boundary layer beneath the disk. The boundary-layer thickness $\delta \approx \sqrt{2\eta/(\rho \Omega)}$ increases as the air density decreases and may eventually span the whole gap beneath the disk. In this limit, the pressure-insensitive scaling predicted by Moffatt is expected to apply. Setting $\delta=R\sin{\theta}$ yields a transition angle  $\theta_{\delta\sim R\theta}=(\eta^2 / (4 g R^3 \rho^2))^{1/3} \approx 0.02$~rad at 0.1 atm \cite{bildsten_viscous_2002}. We do not observe a clear transition to Moffat's $n=1/3$ regime in Fig. \ref{fig:vac}A; however, we do see that the late-time value of $\theta(t)$ is underestimated by Eq.~\ref{dragfric}, suggesting that a pressure-insensitive dissipation mechanism may play a role in the final moments of the disk’s motion. Because the Reynolds number $Re=2\rho\sqrt{gR^3/(\theta\eta^2)}$ is proportional to $\rho$, a crossover to turbulent flow is unlikely at low pressures. Such a transition may nevertheless occur at ambient pressure \cite{bildsten_viscous_2002}. A detailed numerical solution could clarify the structure of the air drag and its evolution in the final stages of motion.

As a final geometric control, we examine the motion of a steel annulus on glass. In contrast to solid disks, the annulus does not exhibit a crossover to $n=4/9$ as shown in Fig.~\ref{fig:vac}B. Instead, rolling friction appears to dominate the dissipation throughout the motion. This absence of a crossover indicates that removing material from the disk center substantially reduces the viscous air drag, consistent with dissipation arising from shear in the air layer beneath a solid disk. These experiments are performed analogously to the disk measurements, except that three $5 \times 7$ checkerboards are attached around the circumference of the ring to avoid obstructing airflow through the center.

Taken together, these results establish viscous air drag as the dominant dissipation mechanism in the final seconds of motion. Despite this late-time importance, rolling resistance still accounts for most of the energy dissipated over the full trajectory. For example, a 450 g disk spins $\sim400\%$ longer on glass than on steel; similar differences occur in sharp-edged vs. filleted disks. Notably, on glass surfaces, $\Phi_{\mathrm{roll}}$ appears to be nearly independent of disk mass, in disagreement with existing models  \cite{easwar_speeding_2002,le_saux_dynamics_2005,leine_experimental_2009,ma_rolling_2014}. This causes heavier disks to spin substantially longer than lighter ones; for example, on glass, a 450 g disk spins about 150\% longer than a 100 g disk. As discussed in the Supplemental Material (Fig. S5 \cite{supplemental}), this trend is noticeable in the final 10~s of motion in Fig. 2A and is consistent with adhesion-mediated rolling resistance \cite{hanrahan_adhesion-dominated_2014}. We do not observe this trend on steel or aluminum, where more strongly load-dependent mechanisms such as plasticity or hysteresis are likely dominant.

The late-time dynamics that we measure here provide quantitative support for a long-standing prediction that air-drag governs the dissipation in the final seconds leading up to the finite-time singularity in Euler's disk \cite{moffatt_eulers_2000, engh_numismatic_2000, bildsten_viscous_2002}. Furthermore, the results hint at a rich rolling-friction regime at early-times that may be adhesion-mediated on glass. Rolling friction remains poorly understood and is often described empirically despite its profound importance in mechanical devices \cite{johnson_contact_1985}. These results therefore motivate systematic studies of the parameter-dependent prefactors governing rolling contact. Owing to its simplicity and repeatability, Euler’s disk provides an attractive platform for such investigations.

\vspace{5mm}
This work was supported by the National Science Foundation of the United States through the Harvard University MRSEC (DMR-2011754) as well as the Wyss Institute for Biologically Inspired Engineering at Harvard University. We thank Tadashi Tokieda for introducing us to Euler's Disk, Lars Bildsten for helpful feedback, as well as Michael Landry and Adam Frim for their contributions as undergraduate researchers.

\bibliography{euler}

\clearpage
\section{Supplementary Material}

\renewcommand{\thefigure}{S\arabic{figure}}
\renewcommand{\thetable}{S\arabic{table}}
\renewcommand{\theequation}{S\arabic{equation}}
\setcounter{figure}{0}
\setcounter{table}{0}
\setcounter{equation}{0}
\subsection{Physical Parameters}

\begin{table}[H]
\centering
\resizebox{\linewidth}{!}{
\begin{tabular}{|l|r|r|r|r|}
\hline
\textbf{Material}                & \multicolumn{1}{l|}{\textbf{Diameter (mm)}} & \multicolumn{1}{l|}{\cellcolor[HTML]{FFFFFF}\textbf{Thickness (mm)}} & \multicolumn{1}{l|}{\textbf{Mass (g)}} & \multicolumn{1}{l|}{\textbf{Fillet Radius (mm)}} \\ \hline
\cellcolor[HTML]{FFFFFF}Aluminum & 101.8                                      & 9.5                                                               & 210                                   & 0                                               \\ \hline
\cellcolor[HTML]{FFFFFF}Aluminum & 101.7                                      & 6.4                                                                & 139                                   & 0                                               \\ \hline
\cellcolor[HTML]{FFFFFF}Aluminum & 101.7                                      & 12.7                                                                & 280                                   & 0                                               \\ \hline
\cellcolor[HTML]{FFFFFF}Aluminum & 50.8                                       & 9.5                                                               & 53                                    & 0                                               \\ \hline
\cellcolor[HTML]{FFFFFF}Aluminum & 50.8                                       & 12.7                                                                & 70                                    & 0                                               \\ \hline
\cellcolor[HTML]{FFFFFF}Aluminum & 50.8                                       & 6.4                                                                & 35                                    & 0                                               \\ \hline
\cellcolor[HTML]{FFFFFF}Steel    & 74.4                                       & 3.0                                                              & 102.2                                 & 1.6                                             \\ \hline
\cellcolor[HTML]{FFFFFF}Steel    & 74.4                                       & 13.6                                                             & 460                                   & 0                                               \\ \hline
\cellcolor[HTML]{FFFFFF}Steel    & 74.4                                       & 7.4                                                              & 252.3                                 & 1.6                                             \\ \hline
\cellcolor[HTML]{FFFFFF}Aluminum & 74.4                                       & 7.4                                                              & 85.1                                  & 1.6                                             \\ \hline
\cellcolor[HTML]{FFFFFF}Aluminum & 74.4                                       & 10.3                                                              & 120.5                                 & 1.6                                             \\ \hline
\cellcolor[HTML]{FFFFFF}Aluminum & 74.4                                       & 3.1                                                              & 36.5                                  & 1.6                                             \\ \hline
\cellcolor[HTML]{FFFFFF}Aluminum & 74.5                                       & 7.3                                                              & 84                                    & 1.6                                             \\ \hline
\cellcolor[HTML]{FFFFFF}Aluminum & 74.4                                       & 15.9                                                             & 186.5                                 & 1.6                                             \\ \hline                                      
\cellcolor[HTML]{FFFFFF}Aluminum & 62.5                                       & 10.2                                                             & 85                                    & 0                                               \\ \hline
\cellcolor[HTML]{FFFFFF}Aluminum & 76.2                                       & 13.4                                                             & 165                                   & 0                                               \\ \hline
\cellcolor[HTML]{FFFFFF}Steel    & 75                                         & 13.0                                                             & 445                                   & 1.6                                            \\ \hline
\end{tabular}
}
\caption{Disk physical parameters.}
\label{table:1}
\end{table}

\renewcommand{\arraystretch}{1}

\begin{table}[H]
\centering
\resizebox{\linewidth}{!}{
\begin{tabular}{|l|r|r|r|r|r|}
\hline
\textbf{Material} &
\multicolumn{1}{c|}{\textbf{\shortstack[c]{\vspace{2pt}\\Outer\\Diameter (mm)}}} &
\multicolumn{1}{c|}{\textbf{\shortstack[c]{Inner\\Diameter (mm)}}} &
\multicolumn{1}{c|}{\textbf{Thickness (mm)}} &
\multicolumn{1}{c|}{\textbf{Mass (g)}} &
\multicolumn{1}{c|}{\textbf{\shortstack[c]{Fillet\\Radius (mm)}}} \\
\hline
Steel & 108 & 70 & 17.7 & 740 & 1.6 \\
\hline
\end{tabular}
}
\caption{Annulus physical parameters.}
\label{table:annulus}
\end{table}

\renewcommand{\arraystretch}{1}

\subsection{The Final Moments of Motion and $t_f$}

We define $t_f$ as the extrapolated terminal time of the power-law regime described by Eq.~3 of the main text. In practice, $t_f$ is chosen such that the final $0.1$~s of the trajectory is maximally linear in log--log space, minimizing residuals of the power-law fit. This choice generally places $t_f$ 2--5~ms before the disk ceases all observable motion. In general, the fitted power-law exponent $n$ is somewhat sensitive to the precise choice of $t_f$, since small shifts in $t_f$ rescale the horizontal axis in log--log space. The prefactor of the power law, however, is comparatively insensitive to such shifts, so the inferred dissipation mechanisms remain robust to reasonable variations in $t_f$.

\begin{figure}[H]
\includegraphics[width=1\textwidth]{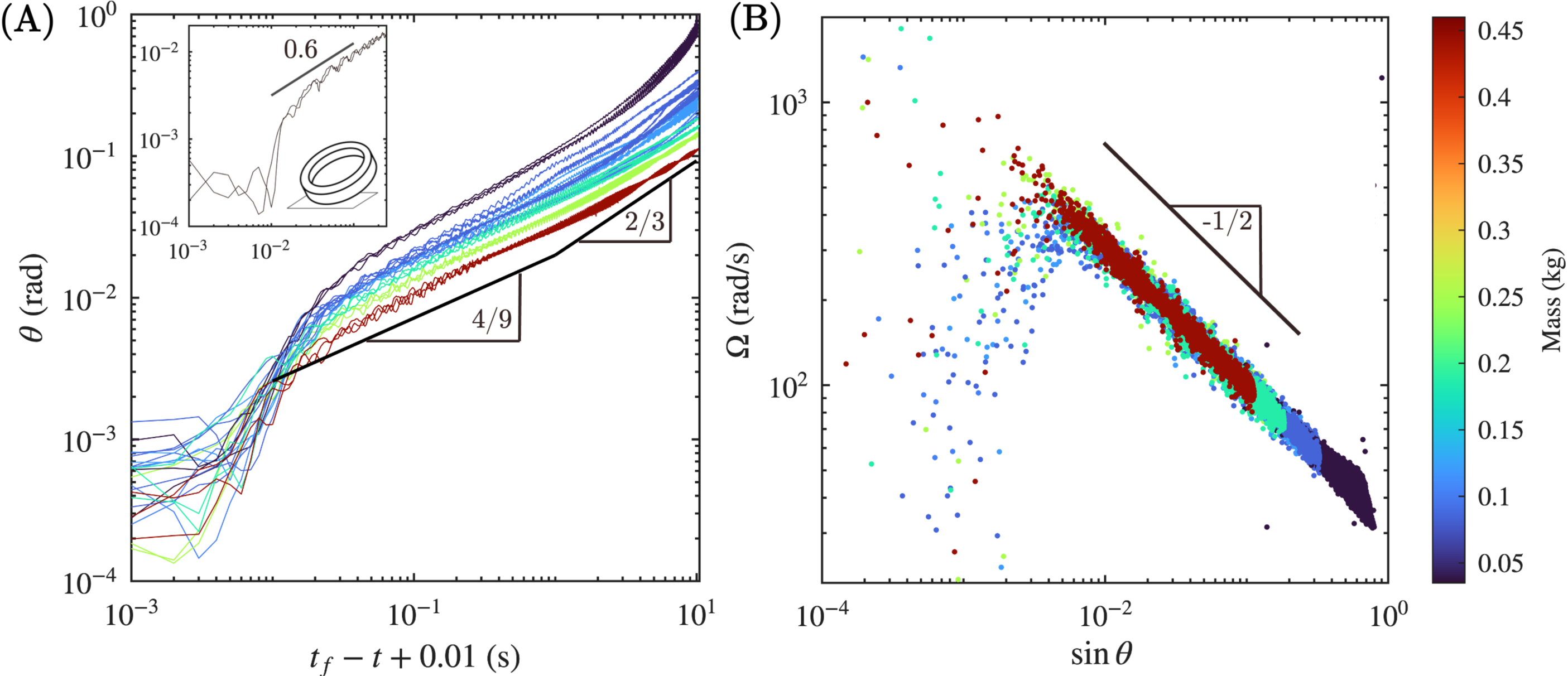}
\caption{\label{fig:extratime} 
\textbf{A)} $\theta$ vs. $t_f-t+0.01$ for filleted disks of varying mass on glass. The data in this graph is translated by $10$~ms to show the motion that occurs after $t_f$. The noise floor for $\theta$ is visible as the leftmost data $\sim 10^{-4}$~rad. Inset: $\theta$ vs. $t_f-t+0.01$ for the steel ring. 
\textbf{B)} Spin frequency $\Omega$ vs. tilt angle $\theta$ for the data shown in \textbf{A}.
}
\end{figure}

To illustrate the dynamics after $t_f$, we plot $\theta$ versus $t_f - t + 0.01$ in Fig.~\ref{fig:extratime}A, which shifts the data by $10$~ms and reveals the residual motion following $t_f$. During this brief interval the disk no longer follows Eq.~1 of the main text, indicating that the assumptions underlying the rolling solution have broken down (Fig.~\ref{fig:extratime}B).

We interpret this brief period of motion after $t_f$ as a loss of contact between the disk and the surface \cite{kessler_ringing_2002,borisov_loss_2015,collins_eulers_2022}. This loss of contact can be predicted by estimating when the normal force between the disk and the surface vanishes. The normal force is approximately
\[
N \approx mg - mR\ddot{\theta},
\]
so contact is lost when $R\ddot{\theta}=g$ \cite{moffatt_eulers_2000}. Using Eq.~4 of the main text, this condition predicts contact loss at
\[
t_f-t \approx 0.68 \frac{R^{8/7}(\eta \rho)^{1/7}}{g^{4/7}m^{2/7}}.
\]

For the heaviest disk ($0.45$~kg) this estimate gives $t_f-t \approx 1.6$~ms, while for the lightest disk ($36$~g) it predicts $t_f-t \approx 3.2$~ms, consistent with the observed 2--5~ms interval of residual motion. Due to noise in the measured trajectories it is difficult to estimate $\ddot{\theta}$ directly from the data without resorting to fitting to a function.

As discussed in Ref.~\cite{collins_eulers_2022}, the thin air layer beneath the disk can slow its final descent toward the surface. Lighter and wider disks are therefore expected to remain separated from the surface for longer times. Consistent with this expectation, we observe a shorter interval of motion after $t_f$ for a steel ring (Fig.~\ref{fig:extratime}A inset). A ring allows air to escape more readily than a disk and therefore falls more quickly.
\clearpage
\subsection{Additional $\theta(t_f-t)$ and $\Omega(\theta)$ Plots}
\vspace{1mm}

\begin{figure}[!htp]
\centering
\includegraphics[width=\textwidth]{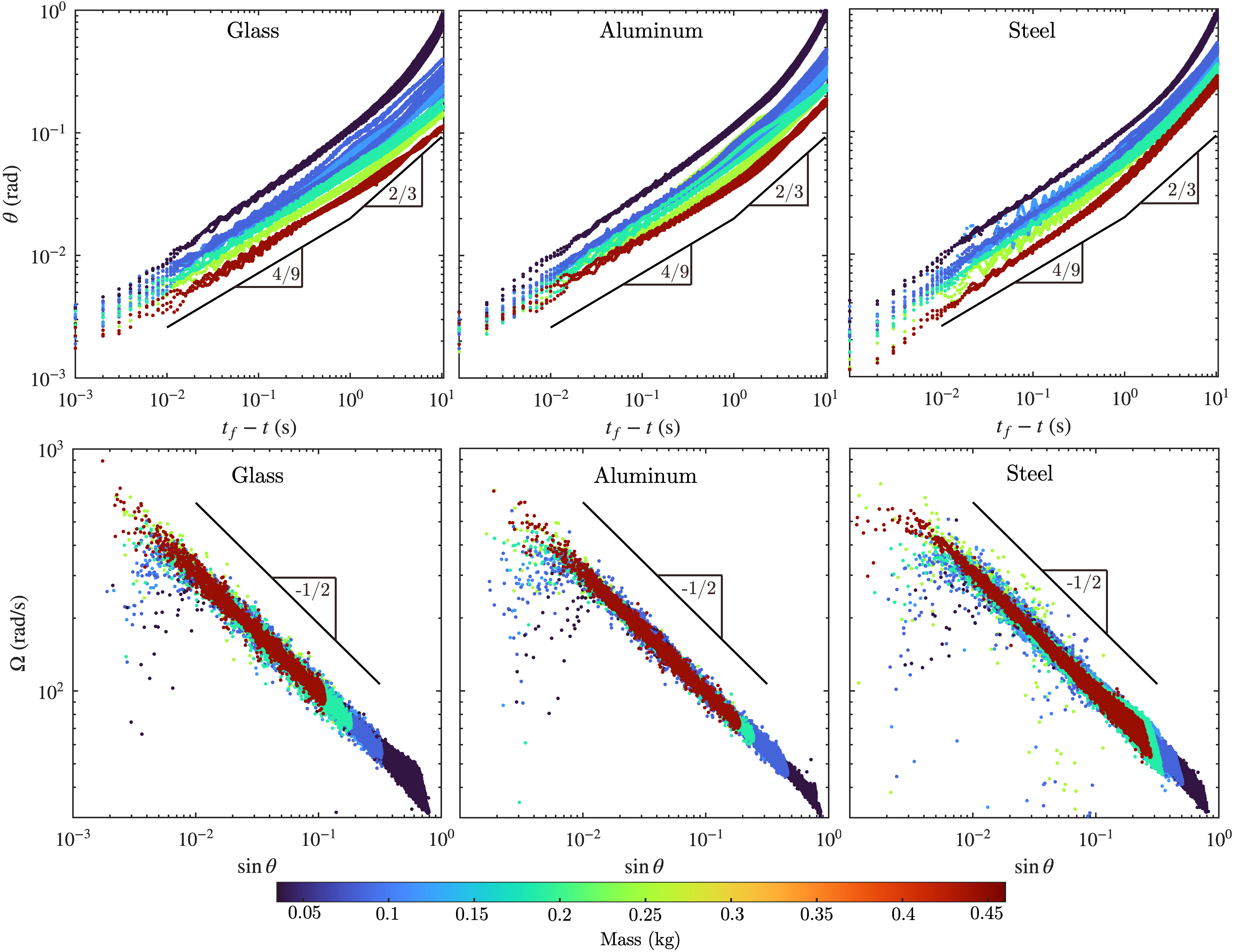}
\caption{\label{fig:fillettedtheta} \textbf{Top:} $\theta$ vs. $t_f-t$ for \textit{filleted} disks of varying mass and constant diameter on glass, aluminum, and steel sheets. Three repetitions per disk. \textbf{Bottom:} Precession frequency $\Omega$ vs. $\sin\theta$ for the data shown above.}
\end{figure}
\vspace{50mm}

\begin{figure}[!htp]
\includegraphics[width=\textwidth]{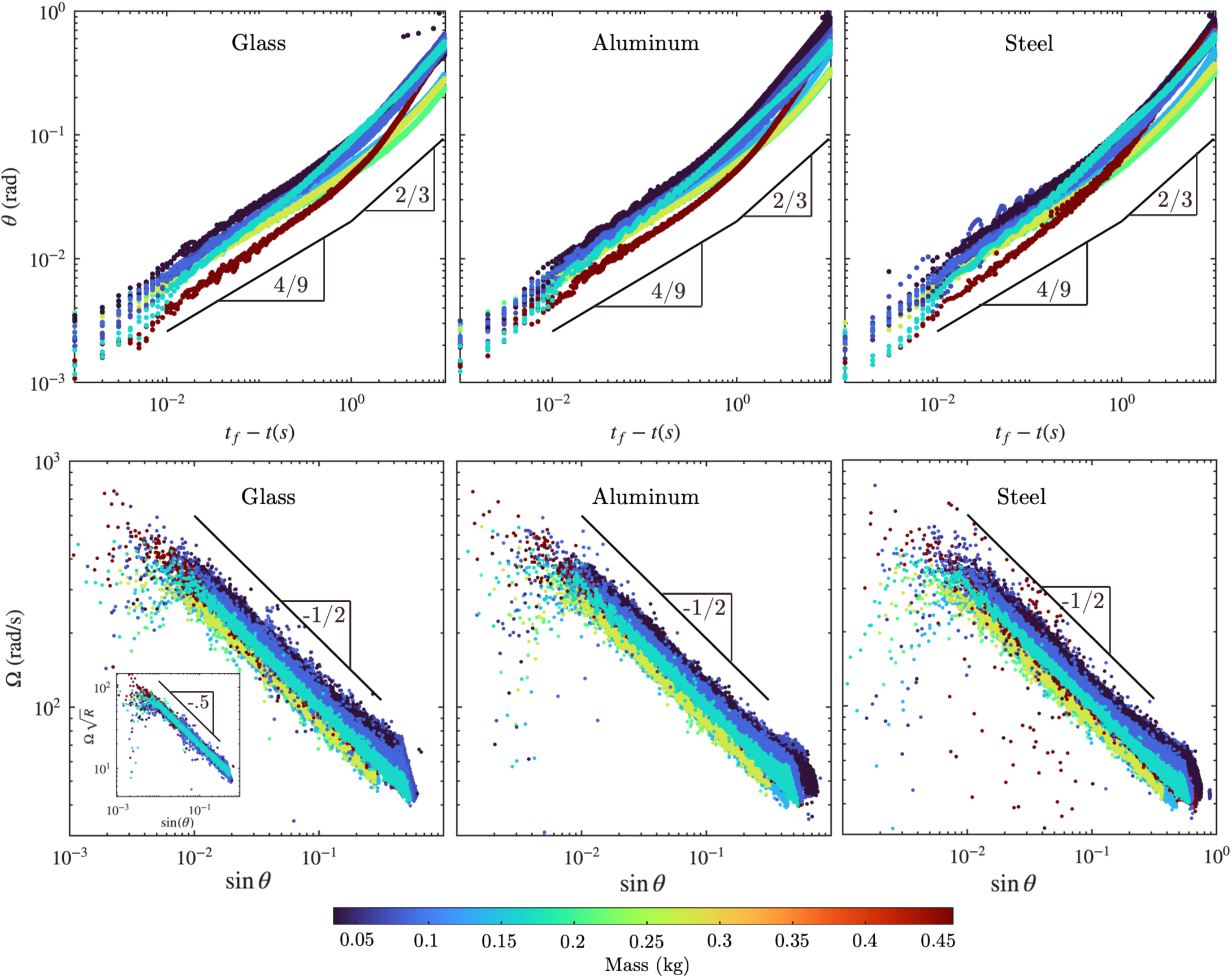}
\caption{\label{fig:sharptheta} \textbf{Top:} $\theta$ vs. $t_f-t$ for \textit{sharp} disks of varying mass and diameter on glass, aluminum, and steel sheets. Three repetitions per disk. \textbf{Bottom:} Precession frequency $\Omega$ vs. $\sin\theta$ for the data shown above. The inset on the left shows that the curves collapse when scaled by $\sqrt{R}$ consistent with Eq. 1 in the main text.}
\end{figure}

\begin{figure}[H]
\includegraphics[width=1\textwidth]{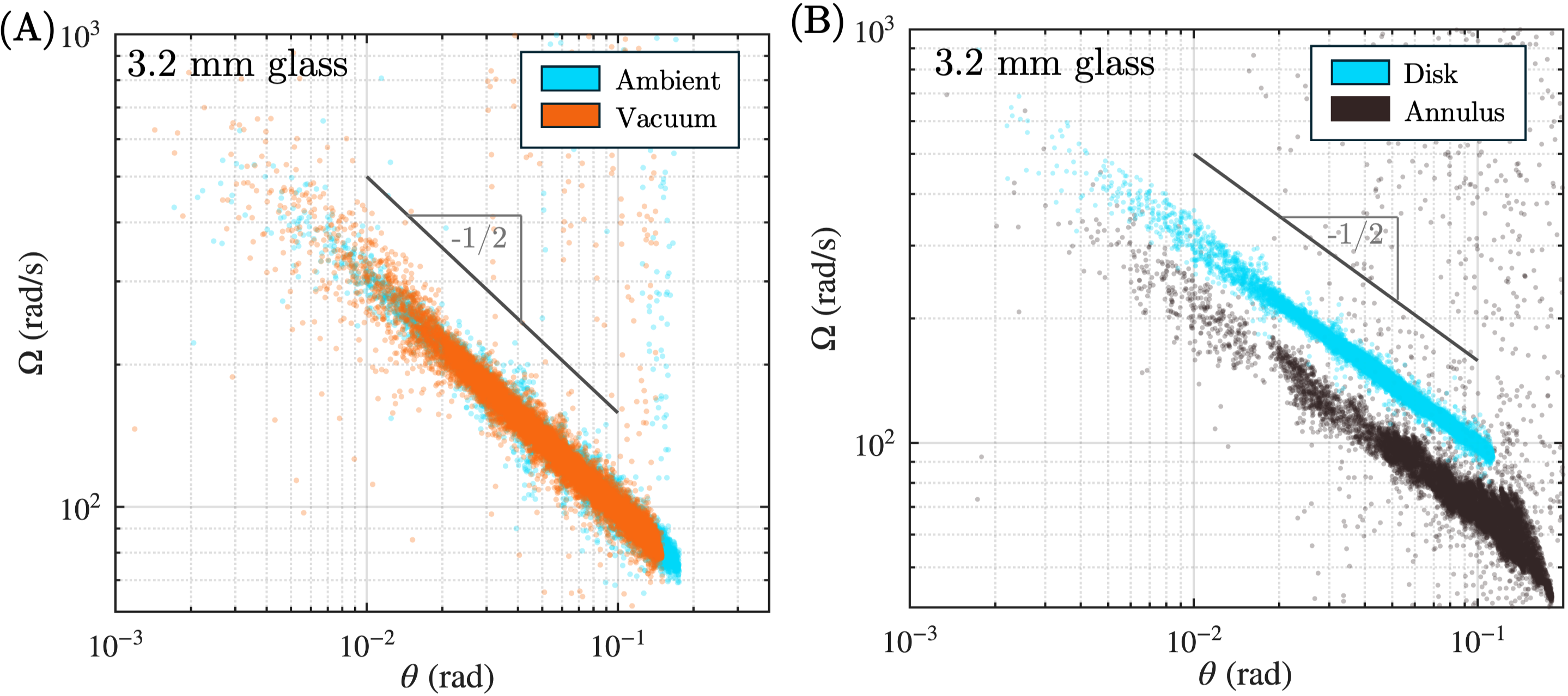}
\caption{\label{fig:omegavac} \textbf{A)} Precession frequency $\Omega$ vs. tilt angle $\theta$ on 3.2 mm glass for a 445 g disk inside and outside of a vacuum. \textbf{B)} Spin frequency $\Omega$ vs. tilt angle $\theta$ for a 445 g disk and for a 740 g annulus on glass.}
\end{figure}

\newpage
\subsection{Mass Dependence of Rolling Friction}
Dissipation due to rolling friction is commonly modeled as the product of a frictional torque and $\Omega$,
\begin{equation}
    \Phi_{\mathrm{roll}} = \mu m g R \cos(\theta) \, \Omega,
    \label{phi_roll}
\end{equation}
where $\mu$ is a coefficient of friction \cite{easwar_speeding_2002,le_saux_dynamics_2005,leine_experimental_2009,ma_rolling_2014}. Integrating this with Eqs. 1 and 2 of the main text yields
\begin{equation}
\theta(t) =\mu^{2/3} \left( \frac{4g}{R} \right)^{1/3}(t_f - t)^{2/3}
\label{roll}
\end{equation}
which is notably independent of the disk mass.

\begin{figure}[H]
\includegraphics[width=0.98\textwidth]{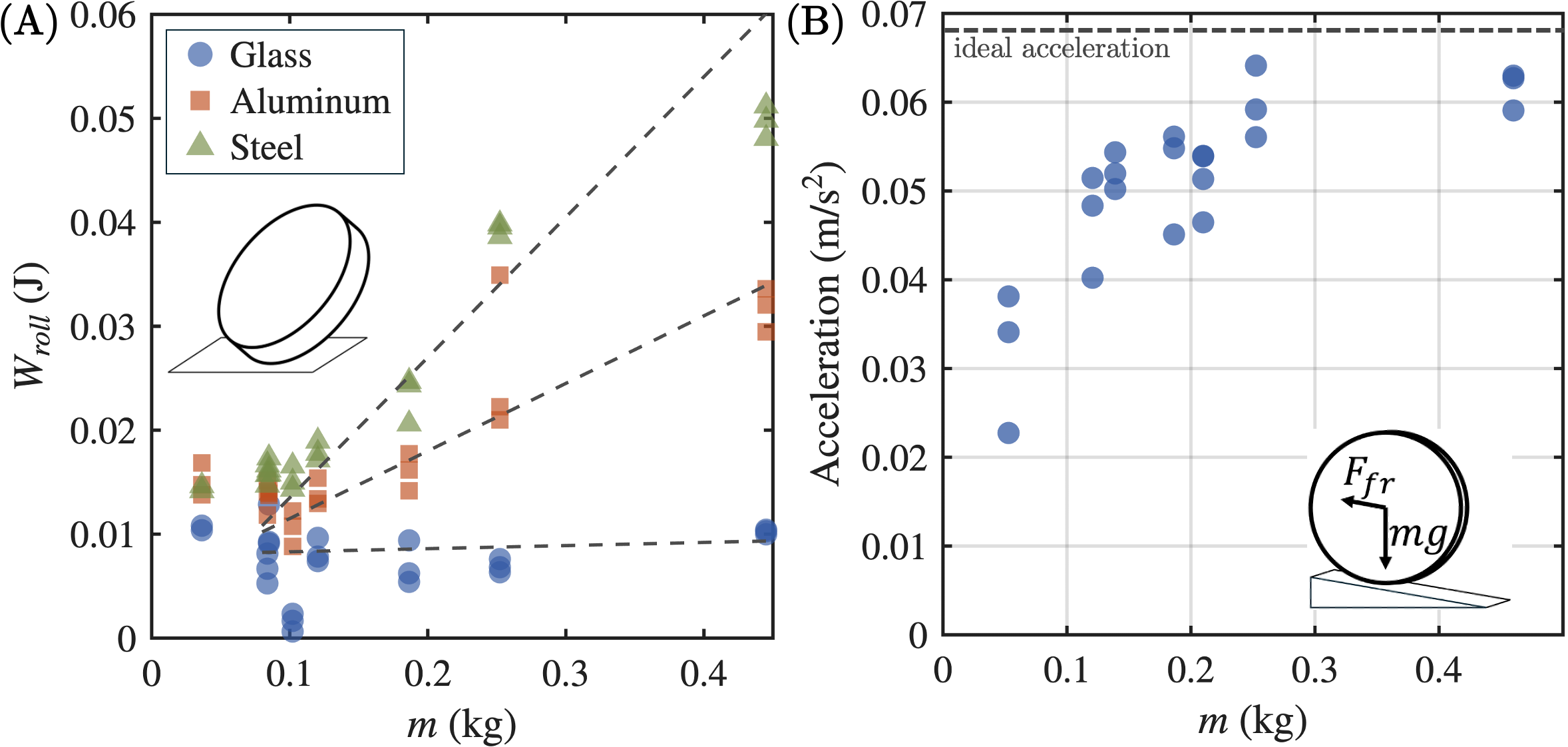}
\caption{\label{fig:adhesion} \textbf{A)} Energy dissipated due to rolling friction $W_{roll}=W_{tot}-W_{drag}$ during the final 10 s of motion for filleted disks. Note that the lightest disks deviate from the trend due to their large $n$ (see Fig. \ref{fig:fillettedtheta}). \textbf{B)} Average acceleration for disks of varying mass rolling down an inclined glass plate. The dotted line indicates the estimated acceleration in the absence of friction $a=\frac{2}{3}g\sin{\alpha}=$ 0.068 m/s$^2$}
\end{figure}

An interesting feature of the early-time dynamics on glass is the unexpected dependence of the rolling-friction on disk mass. Contemporary rolling friction models including Eq. \ref{roll}, are independent of mass, and yet the early-time data in Fig. 2A/S2 still shows a strong gradation with mass that is larger than can be accounted for by air-drag alone. We can estimate the energy dissipated by rolling friction, $W_{\mathrm{roll}}$, over the final 10~s of motion by subtracting the energy dissipated by air drag, $W_{\mathrm{drag}}$, from the total mechanical energy such that
\[
W_{\mathrm{roll}} = \frac{3}{2} m g R \sin(\theta(t_i)) - W_{\mathrm{drag}}.
\]
We obtain the tilt angle at $t_f-t=10$~s, $\theta(t_i)$, from the data in Fig. \ref{fig:fillettedtheta}, and $W_{\mathrm{drag}}$ is calculated by integrating over the data in Fig. \ref{fig:fillettedtheta} such that
\[
W_{\mathrm{drag}} = \int_{t_f-10}^{t_f} 4 g^{5/4} R^{11/4} \sqrt{\eta \rho}\, \theta(t)^{-5/4}\, dt.
\]
$W_{\mathrm{roll}}$ for aluminum and steel increases linearly with disk mass over the range studied, consistent with Eq. \ref{phi_roll}. By contrast, $W_{\mathrm{roll}}$ for glass seems to exhibit no dependence at all on disk mass; see Fig. \ref{fig:adhesion}A. This finding is corroborated by stopping-time measurements. On high-friction surfaces such as plastic or steel, heavier disks spin slightly longer than lighter ones; for example, a 450~g disk typically spins about $20\%$ longer than a 100~g disk, a factor that may be accounted for by air-drag. Conversely, on glass, the same increase in mass leads to an increase in the stopping time of approximately $150\%$.

As an independent comparison, we examine the acceleration of disks rolling on their edge down a glass sheet inclined at an angle of $\alpha=0.6 \pm 0.1^\circ$. Seven disks of varying mass are released from rest at the top of the incline, and their motion is recorded at 240~frames per second. The centroid of the checkerboard points is tracked to obtain the disk velocity as a function of time, and the average acceleration is extracted from a linear fit. In this configuration, heavier disks are observed to accelerate faster than lighter ones, as shown in Fig.~\ref{fig:adhesion}B; this is consistent with the sublinear mass dependence inferred from the disk experiments in Fig. \ref{fig:adhesion}A. The maximum disk speed in these tests is approximately $0.15$~m/s, for which air resistance is expected to be several orders of magnitude smaller than the observed rolling resistance. We observe no systematic dependence of the acceleration on disk radius or thickness. 

Taken together, these results provide an independent indication that rolling friction on glass surfaces exhibits a weak dependence on normal load. 
Rolling friction in general may arise from a combination of mechanisms, including hysteresis, plastic deformation, microslip, micro-impacts, and adhesion \cite{johnson_contact_1985}. Among these, adhesion is expected to exhibit a particularly weak dependence on normal load \cite{hanrahan_adhesion-dominated_2014} and is therefore a plausible contributor to the observed sublinear scaling on glass. For rolling-contact systems with larger loads or rougher surfaces, we expect more strongly load-dependent mechanisms such as hysteresis or plastic deformation to dominate \cite{johnson_contact_1985}.

\end{document}